\RequirePackage{fixltx2e} 
\documentclass[aps,prl,twocolumn,superscriptaddress]{revtex4-1}

\usepackage{amsmath}
\usepackage{textcomp}
\usepackage{amsfonts}
\usepackage{amssymb}
\usepackage{inputenc}
\usepackage{graphicx}
\usepackage{color}
\usepackage{dblfloatfix}
\usepackage{placeins}
\usepackage{natbib}
\usepackage{url}
\usepackage{textcase}
\usepackage{bm}

\begin{document}

\title{Stripe and short range order in the charge density wave of 1\textit{T}-Cu$_{\text{x}}$TiSe$_{\text{2}}$ }

\author{A.M. Novello}

\author{M. Spera}

\author{A. Scarfato}

\author{A. Ubaldini}

\author{E. Giannini}

\affiliation{Department of Quantum Matter Physics, University of Geneva, 24 Quai Ernest-Ansermet, CH-1211 Geneva 4, Switzerland}

\author{D.R. Bowler}
\affiliation{London Centre for Nanotechnology and Department of Physics and Astronomy, University College London,
London WC1E 6BT, United Kingdom}

\author{Ch. Renner}
 \email{Corresponding author. \\ christoph.renner@unige.ch}
 \affiliation{Department of Quantum Matter Physics, University of Geneva, 24 Quai Ernest-Ansermet, CH-1211 Geneva 4, Switzerland}

\date{\today}

\begin{abstract}
We study the impact of Cu intercalation on the charge density wave (CDW) in 1\textit{T}-Cu$_{\text{x}}$TiSe$_{\text{2}}$ by scanning tunneling microscopy and spectroscopy. Cu atoms, identified through density functional theory modeling, are found to intercalate randomly on the octahedral site in the van der Waals gap and to dope delocalized electrons near the Fermi level. While the CDW modulation period does not depend on Cu content, we observe the formation of charge stripe domains at low Cu content (x$<$0.02) and a breaking up of the commensurate order into 2$\times$2 domains at higher Cu content. The latter shrink with increasing Cu concentration and tend to be phase-shifted. These findings invalidate a proposed excitonic pairing as the primary CDW formation mechanism in this material.
\end{abstract}

\pacs{}

\maketitle

Correlated electron systems are prone to develop distinct electronic ground states, such as superconductivity, charge density waves (CDWs), and spin ordered phases. The nature of the interplay between these ground states is the focus of intense research efforts. A CDW is a spatial modulation of the electron density associated with local lattice distortions. CDWs are found in a number of quasi-two-dimensional superconductors, including transition metal dichalcogenides \cite{Wilson1975}, intercalated graphite \cite{Rahnejat2011}, cuprates \cite{Howald2003,Chang2012,Ghiringhelli2012} and pnictides \cite{Rosenthal2014}. Of particular interest, largely driven by the puzzle of high temperature superconductivity, is whether charge order is competing, cooperating or simply coexisting  with superconductivity \cite{Keimer2015}.
The layered transition metal dichalcogenide 1\textit{T}-TiSe$_{\text{2}}$ offers an attractive playground to explore the interplay between these two electronic ground states, thus potentially contributing to resolving similar outstanding questions in cuprate superconductors and other strongly correlated materials. 

1\textit{T}-TiSe$_{\text{2}}$ consists of a stack of van der Waals (vdW) coupled layers allowing in-situ preparation of surfaces ideally suited for scanning probe investigations by cleaving. When cooled below $T_{CDW}\simeq 200$ K, it undergoes a second-order phase transition into a commensurate 2$\times$2$\times$2 CDW superlattice \cite{Wilson1969,DiSalvo1976}. There is currently no consensus on the origin of the CDW in this material. Two possible scenarios are being considered, one based on a purely electronic process characterized by an excitonic instability \cite{Wilson1969}, while the other one involves a Jahn-Teller (JT) distortion \cite{Hughes_1977}. More refined theories also propose a mixture of these two possible contributions, in the so called indirect JT transition \cite{Kidd2002,Wezel2010,Phan2013}. 

1\textit{T}-TiSe$_{\text{2}}$ becomes superconducting when intercalating more than x=0.04 copper into the vdW gap, with a maximum critical temperature $T_{c}=$4.1 K near x=0.08 \cite{Morosan2006}. Transport measurements  \cite{Morosan2006,Wu2007} indicate the CDW is suppressed upon increasing the Cu content which would suggest a competition with superconductivity. A more recent report of an incommensurate CDW above the superconducting dome in pristine crystals under pressure \cite{Joe2014} suggests a more complex scenario, where CDW fluctuations promote superconductivity. Traces of incommensurate CDW patches have also been found in gated TiSe$_{\text{2}}$ thin films \cite{Li2016}. Here, we focus on the effect of Cu intercalation on the CDW in 1\textit{T}-TiSe$_{\text{2}}$ in an effort to contribute to this discussion from an atomic scale structural and spectroscopic perspective.  

The band structure is an important ingredient for understanding the CDW and superconducting phases. According to angular resolved photoemission (ARPES), the dominant contributions to the electronic band structure near the Fermi level are a Ti 3\textit{d} conduction band at the \textit{L} point and a Se 4\textit{p} valence band at the $\Gamma$ point of the Brillouin zone \cite{Monney2010}. At low temperature, ARPES reveals strong band renormalization with a large transfer of spectral weight into backfolded bands pointing at an excitonic ground state driven CDW transition. On Cu intercalated crystals, ARPES shows the Ti 3\textit{d} conduction band to sink below the Fermi level, indicating an electron donor character of Cu. Increasing the Cu content towards the superconducting composition, the backfolding of the Se 4\textit{p} valence band is found to slowly disappear, suggesting competition between CDW order and superconductivity \cite{Qian2007}. However an alternative explanation contends that the latter is a coincidental response to increasing the chemical potential, which suppresses the CDW, and to the enhancement of states at the Fermi level that unltimately favors the emergence of superconductivity \cite{Zhao2007}.

1\textit{T}-Cu$_{\text{x}}$TiSe$_{\text{2}}$ (0$\le$x$\leq$0.07) single crystals were grown by iodine vapour transport of a stoichiometric mixture of Ti and Se sealed in a quartz ampoule under high vacuum. The Cu content x was adjusted by adding an appropriate amount of metallic Cu to the starting materials. Single crystals were obtained after one week at 650$^\circ$C for pristine (x=0) and 830$^\circ$C for Cu intercalated specimen. These temperatures were chosen to limit the amount of Ti self doping \cite{DiSalvo1976,Hildebrand2016}. The single crystals were cleaved in-situ at room temperature prior to the STM/STS measurements (base pressure below $1\times10^{\text{-10}}$ mBar). We used in-situ conditioned PtIr tips and the bias voltage V$_{\text{bias}}$ was applied to the sample. The differential conductance \textit{dI/dV(V)} curves were acquired using a standard lock-in technique with a 5 mV bias modulation at 413.7 Hz.

DFT modeling of intercalated Cu was performed with the plane wave pseudopotential code VASP \cite{Kresse1996,Kresse1993}, version 5.3.5. Projector-augmented waves \cite{Kresse1999} in a 28.04$\times$28.04 \AA$^2$ rhombohedral unit cell were used with the Perdew-Burke-Ernzerhof (PBE) \cite{Perdew1996} exchange correlation functional and plane wave cutoffs of 295 eV. The 1\textit{T}-TiSe$_\text{2}$ surface was modeled with two layers with the bottom Se layer fixed. A Monkhorst-Pack mesh with 1$\times$1$\times$1 and 2$\times$2$\times$1 \textit{k}-points was used to sample the Brillouin zone of the cell, with the finer grid used to check convergence. The parameters gave an energy difference convergence better than 0.01 eV. During structural relaxations, a tolerance of 0.03 eV/\AA     was applied. STM images were generated following the Tersoff-Hamann \cite{Tersoff1983} approach in which the I(V) characteristic measured by STM is proportional to the integrated local density of states (LDOS) of the surface using the BSKAN code \cite{Hofer2003}.

\begin{figure}
 \includegraphics{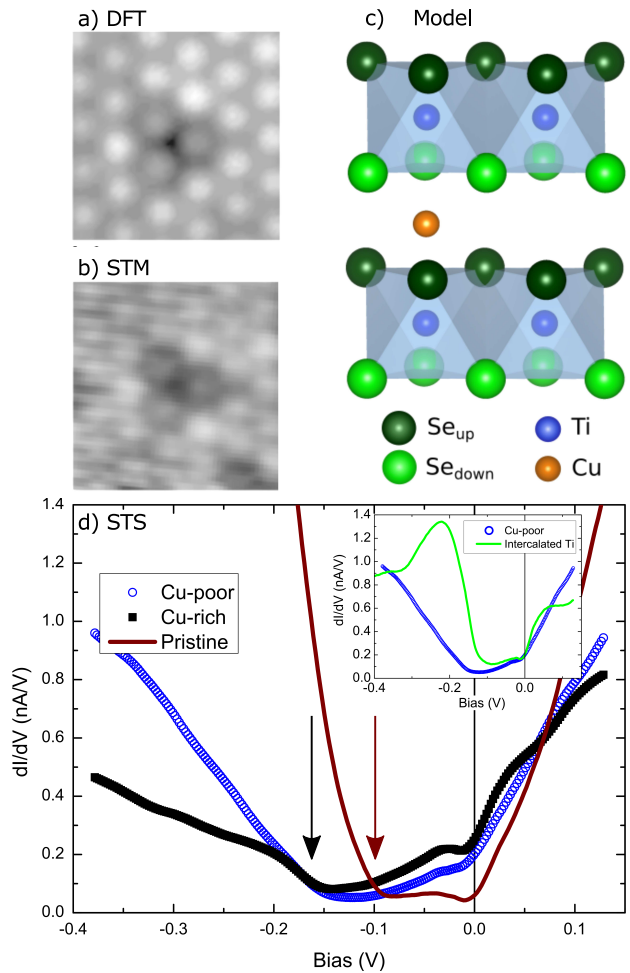}
 \caption{(color online). (a) DFT simulation and (b) STM image of an intercalated Cu atom (1.72 $\times$ 1.72 nm$^{2}$, \textit{V$_\text{bias}$}$\text{ = -1.2 V}$ , \textit{I$_\text{t}$}$\text{ = 30 pA}$). (c) Model of the 1\textit{T}-Cu$_\text{x}$TiSe$_\text{2}$ showing the Cu atom position in the vdW gap. (d) Experimental \textit{dI/dV(V)} curves obtained on 1\textit{T}-TiSe$_\text{2}$ (red) and  1\textit{T}-Cu$_\text{0.012}$TiSe$_\text{2}$ (blue, black) single crystals  (Averaging 20 spectra for each curve, \textit{T} = 1.2 K, \textit{V$_\text{set}$} = 150 mV, \textit{I$_\text{t}$} = 100 pA). Inset: spectra taken in a Cu-free region away from Ti defects (blue) and on an intercalated Ti (green) of the same 1\textit{T}-Cu$_\text{0.012}$TiSe$_\text{2}$ sample.   \label{Fig:1}}
 \end{figure}

A positive identification of the Cu atoms in topographic STM images is paramount to a thorough atomic scale study of the impact of intercalated Cu on the CDW in 1\textit{T}-Cu$_{\text{x}}$TiSe$_{\text{2}}$. While STM imaging readily revealed specific patterns of intercalated Ti and three other dominant single atom defects in 1\textit{T}-TiSe$_\text{2}$ \cite{Hildebrand2014}, the footprint of intercalated Cu proved far more elusive. Guided by DFT modeling, we find that intercalated Cu atoms are only resolved with atomic resolution in a reduced energy window around -1.2 V, with a perfect correspondence between model and data (Fig.\ref{Fig:1}a,b). This is the first solid experimental evidence that intercalated Cu is indeed sitting on the octahedral site in the vdW gap (Fig.\ref{Fig:1}c), as assumed in literature \cite{Titov2001,Stoltz2005}.

\begin{figure}[h]
 \includegraphics{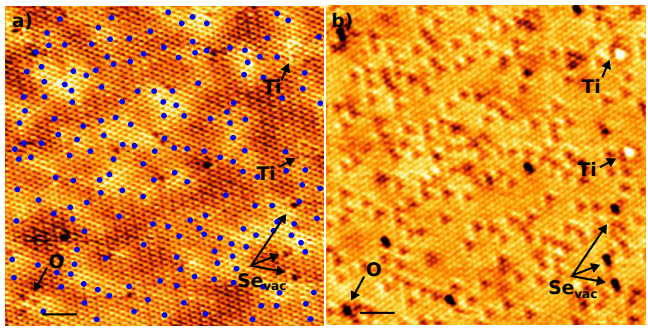}
 \caption{\label{Fig:2} (color online). STM micrographs of the same surface region of 1\textit{T}-Cu$_{\text{0.07}}$TiSe$_{\text{2}}$ at 1.2 K. The set point (\textit{V$_\text{bias}$}$\text{ = 150 mV}$, \textit{I$_\text{t}$} = 30 pA) in (a) is chosen to show the CDW modulation and (-1.2 V, 30 pA) in (b) to show the intercalated Cu atoms. Blue dots represent the Cu atom positions in the vdW gap and labeled arrows point at single atom defects \cite{Hildebrand2014}. Scale bars: 2 nm.}
 \end{figure}

Substantial charge inhomogeneities are seen in STM micrographs of Cu intercalated 1\textit{T}-TiSe$_{\text{2}}$ (Fig.\ref{Fig:2}a), especially at low bias. Imaging the exact same region at $\text{-1.2 V}$ (Fig.\ref{Fig:2}b) where Cu atoms can be resolved one by one, we show that these inhomogeneities are directly linked to intercalated Cu, with bright regions corresponding to an accumulation of Cu atoms. The two distinct STM images in Fig.\ref{Fig:2} have been aligned with atomic scale precision based on single atom defects identified in previous studies \cite{Hildebrand2014, Novello2015}. Such precise identification of single atom defects resolved by STM permits exquisite insight into the microscopic nature of the material and the CDW. The atomic patterns of the defects associated with single atom O, I, Ti and Cu defects imply that the observed CDW modulation resides on the Se sites. The relative orientation of these mostly triangular features further confirm the local 1\textit{T}-polytype of the single crystals investigated. Intercalated Ti was shown to show two different topographic patterns depending on its relative position with respect to the CDW superlattice \cite{Hildebrand2014}. In the case of Cu, no topographic difference is seen instead. Finally, the local Cu concentration of the region under the STM tip can be assessed simply by counting the Cu atoms: the $\sim$200 Cu atoms found within the 18$\times$18 nm$^\text{2}$ of Fig.\ref{Fig:2} are in excellent agreement with the nominal doping x=0.07 of that crystal. 

\begin{figure}
 \includegraphics{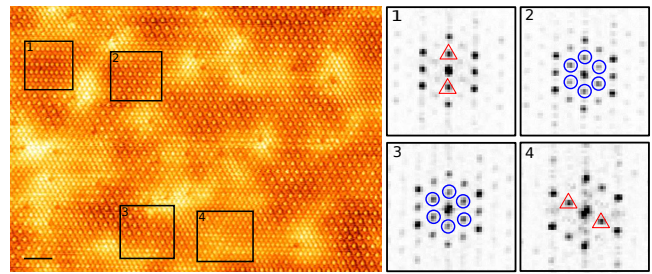}
 \caption{\label{Fig:3} (color online). STM micrograph (\textit{V$_\text{bias}$} = -50 mV, \textit{I$_\text{t}$} = 200 pA) of 1\textit{T}-Cu$_{\text{0.01}}$TiSe$_{\text{2}}$ at 78 K showing 3\textit{Q} CDW (panels 2 and 3) and 1\textit{Q} stripe (panels 1 and 4) domains on an inhomogeneous background due to intercalated Cu atoms. The numbered panels show the regions where the Fourier transforms were calculated, with circles and triangles corresponding to the $q_{3Q}$ and $q_{1Q}$ components, repectively. Scale bar: 2 nm.}
 \end{figure}

Low temperature tunneling spectroscopy of 1\textit{T}-Cu$_\text{x}$TiSe$_{\text{2}}$ (Fig.\ref{Fig:1}d) reveals a gap $\Delta_{\text{CDW}}\simeq$ 80 meV opening below the chemical potential as the material is cooled through $T_{CDW}$ for all Cu concentrations considered. This marked asymmetry and line shape are unusual for a CDW. Note that the gap opening below the Fermi level is consistent with the observed real space charge modulation residing primarily on the Se sites and Se 4\textit{p} states being the main contributions to the valence band. The gap edge in the occupied LDOS is shifting to lower energies with increasing Cu content (arrows in Fig.\ref{Fig:1}d) while the gap is partially filling with states added near the Fermi level. These spectral changes correspond to increasing the chemical potential consistent with the electron donor nature of intercalated Cu. They are sensitive to the local Cu concentration with a higher LDOS measured at $E_F$ in the brighter Cu-rich regions (Fig.\ref{Fig:2}a and Fig.\ref{Fig:3})). These observations are in agreement with previous STM \cite{Iavarone2012} and ARPES \cite{Qian2007,Zhao2007} studies. Here, we demonstrate a direct link between the Cu content and the observed band shift. The origin of the weak singularity at $E_F$, also observed in point-contact tunneling \cite{Luna2015}, remains to be firmly established. It is most likely not a real DOS feature but a manifestation of electron-electron interaction and impurity scattering expected in disordered metals \cite{al_1979}, with disorder due to intercalated Cu and Ti. 

The charge order landscape imaged by STM in 1\textit{T}-Cu$_\text{x}$TiSe$_{\text{2}}$ depends on the Cu content x. The most striking features in filled state STM micrographs of low doped crystals (x$<$0.02) are the appearance of symmetry breaking stripe CDW (1\textit{Q}) domains and, in Cu-rich regions, a reduced amplitude of the 2$\times$2 CDW (3\textit{Q}) (Fig.\ref{Fig:3}a). Tip artifacts are excluded to explain the stripe domains: 1\textit{Q} and 3\textit{Q} charge order is seen along the same scan line - hence with the same tip configuration. We do not measure any systematic spectroscopic difference between the two domains. Stripe charge order has been observed by STM in other layered compounds \cite{Soumyanarayanan2013,Rahnejat2011,Bando1997} and was attributed to local strain. Strain could play a role here as a consequence of the reported 1\textit{T}-TiSe$_{\text{2}}$ unit cell expansion when intercalating Cu \cite{Shkvarin2016}. 

We observe 1\textit{Q} and 3\textit{Q} phases coexisting down to $\text{1.2 K}$ in low doped crystals. They are always commensurate and in perfect registry with the Se lattice. The Fourier transforms show the $q_{1Q}$ component of the stripe phase to be identical to one of the three $q_{3Q}$ components of the 3\textit{Q} phase (Fig.\ref{Fig:3}). This is different from the observations in 2\textit{H}-NbSe$_{\text{2}}$ where they were found to differ by as much as 13\% \cite{Soumyanarayanan2013}. The independence of the 1\textit{Q} and 3\textit{Q} Fourier components on Cu concentration and the marked electron-hole asymmetry of the CDW gap exclude a Fermi surface nesting mechanism to explain the CDW. Fig.\ref{Fig:3} further questions the chiral nature of the CDW in 1\textit{T}-TiSe$_{\text{2}}$ claimed in several STM studies  \cite{Ishioka2010,Iavarone2012} based on a systematic hierarchy in the amplitudes of the $q_{3Q}$ CDW Fourier components. There is no supporting evidence for a chiral CDW in our images. We see no significant amplitude differences in the Fourier components, and only a single $q_{1Q}$ component is present in the stripe domains.

\begin{figure}
 \includegraphics{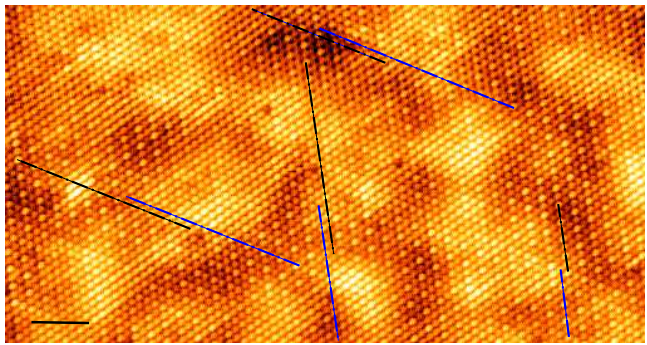}
 \caption{\label{Fig:4} (color online). STM micrograph (\textit{V$_\text{bias}$}$ \text{ = 10 mV}$, \textit{I$_\text{t}$}$\text{ = 60 pA}$) of 1\textit{T}-Cu$_\text{0.07}$TiSe$_\text{2}$ at 1.2 K. The dashed blue/black lines highlight the $\pi$-phase shift between neighbouring CDW domains. Scale bar = 2 nm.}
 \end{figure}

In high doped crystals (x$>$0.05), we do not observe well developed stripe domains but we find the 3\textit{Q} phase to break up into short range ordered nanometer-scale domains (Fig.\ref{Fig:4}). These domains are predominantly surrounded by Cu-rich regions, and tend to be $\pi$-phase shifted with a few atomically sharp boundaries separating adjacent domains. A similar break up of the CDW into phase-shifted domains has been reported recently in Ti intercalated crystals \cite{Hildebrand2016} where the effect is observed already for 2\% Ti content. In both cases, the 2$\times$2 domain size is shrinking with increasing number of intercalated atoms and the charge modulation period is independent of x.

Comparing the impact of Cu and Ti on the charge order and LDOS provides insightful clues on the CDW formation mechanism. Modeling the STM fingerprints of Cu (Fig.\ref{Fig:1}) and Ti \cite{Hildebrand2014} shows they intercalate on the same lattice site in the vdW gap of 1\textit{T}-TiSe$_{\text{2}}$. Both contribute electrons to the system. Ti is a local dopant with a broad localized state below the Fermi level in the tunneling spectra and no apparent shift in chemical potential (inset of Fig.\ref{Fig:1}d). Cu on the other hand is a band dopant inducing a significant shift in the chemical potential (Fig.\ref{Fig:1}d). The peak in the resistivity as a function of temperature at the CDW phase transition is progressively reduced with increasing Cu content while it collapses abruptly for the highest Ti content \cite{Hildebrand2016}. These results can be understood within the exciton CDW formation mechanism. We have shown that increasing Cu gradually shifts the chemical potential up in energy (Fig.\ref{Fig:1}d) and the Ti 3d band deeper below the Fermi level. The system thereby becomes metallic (see also \cite{Morosan2006}) and the formation of excitons is progressively suppressed. Ti doping on the other hand does not shift the chemical potential, leaving the excitonic pairing unchanged until the CDW domain size becomes smaller than the exciton Bohr radius thus abruptly suppressing their formation \cite{Hildebrand2016}. 

However, the exciton scenario is not compatible with the persistence of a clear local charge order observed by STM in crystals where the CDW phase transition is no longer detected by transport. Thus, excitons may enhance long-range CDW correlations resulting in a signal for non-local probes such as transport and ARPES, but the microscopic CDW formation must involve other contributions such as phonons, for example.  A similar picture was proposed in a pump-probe study of 1\textit{T}-TiSe$_{\text{2}}$ \cite{Porer2014}, where it was shown that the lattice distortion is still present when the excitonic order is quenched. Van Wezel \textit{et al.} \cite{Wezel2010} also concluded in a theoretical study that the presence of excitons significantly enhances the CDW order induced by electron phonon coupling. In addition, the gap opening below the Fermi level as measured by tunneling spectroscopy renders a purely electronic mechanism energetically unfavorable.

Finally, and although superconductivity is not the main focus of this work, we note that the persistence of well developed CDW domains up to the highest Cu contents considered suggests that the suppression of the CDW is not the key factor in the emergence of superconductivity in 1\textit{T}-Cu$_\text{x}$TiSe$_\text{2}$. Instead, as already proposed in high-pressure \cite{Joe2014} and carrier injection \cite{Li2016} experiments, superconductivity may arise from the inhomogeneities of the CDW pattern. Moreover, spectroscopy shows Cu intercalation to shift the chemical potential to higher energies and increase the LDOS at the Fermi level providing a favorable electronic configuration for the appearance of superconductivity above x=0.04.

In summary, the unambiguous identification of intercalated Cu atoms in 1\textit{T}-TiSe$_\text{2}$ enables us to specify their precise atomic position in the octahedral site of the vdW gap - unknown so far - and to establish a firm link between Cu doping and STM topographic and spectroscopic features. In particular, we confirm Cu is a donor contributing delocalized electrons at the Fermi level and shifting the chemical potential up in energy in agreement with previous ARPES results \cite{Qian2007,Zhao2007}. We observe a striking instability towards the formation of charge stripes at low Cu concentration (x$<$0.02) and the formation of short-range ordered domains which tend to be phase-shifted at higher Cu contents (x$>$0.05). The charge order period observed by STM imaging is independent on the nominal Cu concentration and persists locally to doping levels where transport and other spectroscopy techniques no longer detect a CDW phase. The present STM/STS study shows that excitonic pairing alone cannot account for the CDW formation and that another, not purely electronic microscopic mechanism must be at play.

This project was supported by the Swiss National Science Foundation through Div.II (grants 144139 and 162517). DFT calculations were performed using the LCN Computing Cluster. We acknowledge stimulating discussions with P. Aebi, B. Hildebrand, Ch. Berthod, J. van Wezel and J. Lorenzana. We thank C. Barreteau for her help with characterizing the single crystals via transport measurements and G. Manfrini and A. Guipet for their skillful technical assistance.

\bibliography{biblio1}{}
\bibliographystyle{apsrev4-1}

\end{document}